\documentclass[%
 aip,jcp,
 amsmath,amssymb,
 reprint,
]{revtex4-1}

\usepackage{graphicx}
\usepackage{dcolumn}
\usepackage{bm}

\usepackage[utf8]{inputenc}
\usepackage[T1]{fontenc}
\usepackage{mathptmx}
\usepackage{listings}
\usepackage{rotating} 

\usepackage{color}
\usepackage{dcolumn} 
\usepackage{bm} 
\usepackage{graphicx}
\usepackage{multirow} 
\usepackage{pifont} 
\usepackage{epsfig}
\usepackage{amsmath} 
\usepackage{subfigure}
\usepackage{float}
\usepackage{booktabs}
\usepackage{tabularx}
\usepackage{natbib}
\usepackage{gensymb}
\usepackage{rotating}
\usepackage[normalem]{ulem}

\begin{document}

\author{Run R. Li}
\affiliation{
             Department of Chemistry and Biochemistry,
             Florida State University,
             Tallahassee, FL 32306-4390}
             
\author{Marcus D. Liebenthal}
\affiliation{
             Department of Chemistry and Biochemistry,
             Florida State University,
             Tallahassee, FL 32306-4390}
             
\author{A. Eugene DePrince III}
\email{adeprince@fsu.edu}
\affiliation{
             Department of Chemistry and Biochemistry,
             Florida State University,
             Tallahassee, FL 32306-4390}

\title{Challenges for variational reduced-density-matrix theory with three-particle $N$-representability conditions}

\begin{abstract}
The direct variational optimization of the two-electron reduced density matrix (2RDM) can provide a reference-independent description of the electronic structure of many-electron systems that naturally captures strong or nondynamic correlation effects. Such variational 2RDM approaches can often provide a highly accurate description of strong electron correlation, provided that the 2RDMs satisfy at least partial three-particle $N$-representability conditions ({\em e.g.}, the T2 condition). However, recent benchmark calculations on hydrogen clusters [J. Chem. Phys. {\bf 153}, 104108 (2020)] suggest that even the T2 condition leads to unacceptably inaccurate results in the case of 2- and 3-dimensional clusters.  We demonstrate that these failures persist under the application of full three-particle $N$-representability conditions (3POS). A variety of correlation metrics are explored in order to identify regimes under which 3POS calculations become unreliable, and we find that the relative squared magnitudes of the cumulant three- and two-particle reduced density matrices correlates reasonably well with the energy error in these systems. However, calculations on other molecular systems reveal that this metric is not a universal indicator for the reliability of reduced-density-matrix theory with 3POS conditions.

\end{abstract}

\maketitle

\section{Introduction}

The efficient and accurate description of large numbers of strongly correlated electrons represents a serious theoretical and computational challenge. The standard zeroth-order approach to this problem, the complete active space self-consistent field (CASSCF)\cite{Roos:1980:157,Siegbahn:1980:323,Siegbahn:1981:2384,Roos:1987:399}  method, becomes impractical for large numbers ({\em i.e.}, $>20$)\cite{Vogiatzis:2017:184111} of active electrons because the complexity of the CASSCF wave function increases exponentially with this number. Consequently, a large number of approximate CASSCF schemes have been devised that are based on CI, including incomplete active space ({\em e.g.}, restricted active space [RAS] ,\cite{Olsen:1988:2185,Malmqvist:1990:5477} generalized active space [GAS] ,\cite{Fleig:2001:4775,Ma:2011:044128,Manni:2013:3375} and related methods\cite{Ivanic:2003:9364,Khait:2004:210}),
stochastic,\cite{Thomas:2015:5316,Manni:2016:1245} and adaptive\cite{HeadGordon20_2340,Evangelista16_161106} CI approaches.
Additional schemes utilize matrix product state representations of the wave function,\cite{Ghosh:2008:144117,Yanai:2009:2178,Wouters:2014:1501,Sun:2017:291,Ma:2017:2533} or reduced density matrices (RDMs).\cite{Mazziotti08_134108,DePrince16_2260} The latter class of methods is unique in that correlated electronic structure is described solely by RDMs, without invoking any sort of wave function expansion. This choice can be desirable, as RDM-based methods can describe complex systems with an effort that formally increases only polynomially with system size.

It is well known that the $N$-electron wave function contains more information than is required to evaluate the electronic energy. Because the electronic Hamiltonian contains at most pairwise interactions, the electronic energy can be expressed exactly in terms of the two-electron RDM (2RDM), and the 2RDM is thus a natural descriptor of the electronic structure.  The elements of the 2RDM could, in principle, be determined directly by minimizing the energy with respect to their variations, provided that one could guarantee that the optimal 2RDM is derivable from an antisymmetrized $N$-electron wave function (or an ensemble of such wave functions). Such a 2RDM is said to be $N$-representable.\cite{Coleman63_668} In the variational 2RDM (v2RDM) approach, \cite{Husimi:1940:264, Lowdin:1955:1474,Mayer:1955:1579,Percus64_1756,Rosina75_868,Rosina75_221,Garrod75_300,Rosina79_1366,Erdahl79_147,Fujisawa01_8282,Mazziotti02_062511,Mazziotti06_032501,Percus04_2095,Zhao07_553,Lewin06_064101,Bultinck09_032508,DePrince16_423,DeBaerdemacker11_1235,VanNeck15_4064,DeBaerdemacker18_024105,Mazziotti17_084101,Ayers09_5558,Bultinck10_114113,Cooper11_054115,DePrince19_032509,Mazziotti08_134108,DePrince16_2260,Mazziotti16_153001} the energy is minimized subject to the elements of the 2RDM while enforcing necessary $N$-representability conditions, which can be organized according to particle number.\cite{Erdahl01_042113,Mazziotti12_263002} For example, at the two-particle level, we have the PQG conditions of Garrod and Percus,\cite{Percus64_1756} while the so-called T2 condition\cite{Erdahl78_697,Percus04_2095} is classified as a partial three-particle condition. The v2RDM approach is systematically improvable in the sense that, as additional higher-order conditions are applied, the v2RDM energy approaches the full CI one. 

The optimization of the 2RDM subject to a set of $N$-representability conditions is a semidefinite programming (SDP) problem. With an appropriate SDP algorithm,\cite{Mazziotti04_213001,Mazziotti11_083001} the v2RDM approach with PQG conditions can be realized at $\mathcal{O}(n^6)$ floating-point cost, where $n$ is the number of active orbitals. This low scaling allows for the efficient description of large numbers of strongly-correlated electrons -- up to 64 electrons distributed among 64 orbitals,\cite{DePrince19_6164} for example. Hence, the v2RDM approach affords the ability to make qualitative predictions of electronic structure in large systems with large active spaces, {\em e.g.}, allowing one to characterize the emergence of polyradical character in one-\cite{Mazziotti08_134108,DePrince16_423} and two-dimensional\cite{Mazziotti11_5632,DePrince19_6164} graphene nanoribbons. Nevertheless, it is well-known that the v2RDM approach with two-particle constraints often significantly over-correlates electrons (by as much as 20\% in simple diatomic molecules at equilibrium\cite{Mazziotti_11_052506}), whereas partial three-particle conditions ({e.g.}, T2) and full three-particle conditions (3POS) provide significantly more accurate results, albeit at higher [$\mathcal{O}(n^9)$] floating-point cost. Indeed, for small molecules described by small basis sets, v2RDM calculations with 3POS conditions seldom result in deviations from full CI energies that are larger than $1\times 10^{-3}$ E$_{\rm h}$, even at far-from-equilibrium geometries where non-dynamic correlation effects become important.\cite{Mazziotti06_032501} 

Recent benchmark calculations on hydrogen clusters\cite{Evangelista20_104018} have supplied additional evidence that the v2RDM approach provides a reasonable description of strongly correlated systems where many other conventional approaches fail, particularly when enforcing partial three-particle constraints. However, Ref.~\citenum{Evangelista20_104018} also revealed unacceptably large errors in PQG+T2 calculations in certain cases, with errors on the order of 10-50 mE$_{\rm h}$ for 2- and 3-dimensional hydrogen clusters at stretched geometries. These results call into question the notion that the application of (partial) three-particle $N$-representability conditions generally results in a quantitatively correct description of electronic structure. In light of these observations, we assess herein the quality of v2RDM calculations that enforce full three-particle $N$-representability conditions for these problematic cases, and we find that 3POS, too, sometimes yields unacceptably inaccurate results. We find that the relative importance of three- and two-body correlations, as measured by the relative squared magnitudes of the cumulant three- and two-particle RDMs, provides a useful gauge for the reliability of v2RDM calculations that enforce 3POS conditions in these systems. Additional calculations on small-molecule systems, however, suggest that this metric is not a universal indicator of the reliability of 3POS.

\section{Theory}

\label{SEC:THEORY}

The non-relativistic electronic energy for a many-electron system may be be expressed as
\begin{equation}
    \label{EQN:energy}
        E =   \frac{1}{2} \sum_{\sigma\tau}\sum_{pqrs} {}^2D^{p_{\sigma}q_{\tau}}_{r_{\sigma}s_{\tau}} ( pr | qs ) + \sum_{\sigma} \sum_{pq} {}^1D^{p_{\sigma}}_{q_\sigma} h_{pq},
\end{equation}
where $(pr|qs)$ represents an electric repulsion integral in Mulliken notation, $h_{pq}$ represents the sum of electron kinetic energy and electron--nucleus potential energy integrals, the labels $p$, $q$, $r$, and $s$ refer to orthonormal spatial orbitals, and {\color{black}$\sigma$ and $\tau$} indicate $\alpha$- or $\beta$-spin functions. The symbols ${}^1D^{p_{\sigma}}_{q_\sigma}$ and ${}^2D^{p_{\sigma}q_{\tau}}_{r_{\sigma}s_{\tau}}$ refer to elements of the one- and two-particle reduced density matrices (the 1RDM and 2RDM), which are defined as
\begin{equation}
    \label{EQN:D1}
        {}^1D^{p_\sigma}_{q_\sigma} =
            \langle \Psi |\hat{a}^\dagger_{p \sigma} \hat{a}_{q \sigma}| \Psi \rangle.
\end{equation}
and
\begin{equation}
    \label{EQN:D2}
        {}^2D^{p_\sigma q_\tau}_{r_\sigma s_\tau} =
              \langle \Psi |
              \hat{a}^\dagger_{p \sigma}  \hat{a}^\dagger_{q \tau} \hat{a}_{s \tau} \hat{a}_{r \sigma}
              | \Psi \rangle,
\end{equation}
respectively. Here, $\hat{a}^\dagger$ and $\hat{a}$ represent the fermionic creation and annihilation operators of second quantization, respectively. The electronic energy given by Eq.~\ref{EQN:energy} is an exact functional of the 1RDM and the 2RDM, which leads the tantalizing prospect of the direct optimization of the the elements of these matrices, without knowledge of the wave function. However, physically meaningful results depend upon our ability to guarantee that the RDMs are derivable from an ensemble of antisymmetrized $N$-electron wave functions. In the following subsections, we review some necessary yet insufficient conditions for the ensemble $N$-representability of the 2RDM that can be defined in terms RDMs involving up to three creation/annihilation operator pairs. 

\subsection{Two-particle $N$-representability conditions}

We begin by defining basic statistical conditions that must be satisfied by any RDM representing a many-fermion system.  First,  RDMs must be Hermitian, and they must respect particle exchange symmetry. For example, the 1RDM must satisfy ${}^1D^{p_\sigma}_{q_\sigma} = {}^1D_{p_\sigma}^{q_\sigma}$, while the 2RDM must satisfy 
    ${}^2D^{p_\sigma q_\tau}_{r_\sigma s_\tau} = {}^2D^{r_\sigma s_\tau}_{p_\sigma q_\tau}$, 
and
\begin{equation}
    \label{EQN:PARTICLE_EXCHANGE}
     {}^2D^{p_\sigma q_\tau}_{r_\sigma s_\tau} = -{}^2D^{p_\sigma q_\tau}_{s_\tau r_\sigma}=- {}^2D^{q_\tau p_\sigma}_{r_\sigma s_\tau}= {}^2D^{q_\tau p_\sigma}_{s_\tau r_\sigma}
\end{equation}
Second, because the electronic Hamiltonian does not break particle-number symmetry, the wave function must be an eigenfunction of the particle number operator. As a result, the RDMs must satisfy trace constraints of the form
\begin{eqnarray}
     \label{EQN:TRACE1}
     \langle \Psi |\hat{N_\sigma}| \Psi \rangle &=& N_\sigma,\\
          \label{EQN:TRACE2}
     \langle \Psi |\hat{N_\sigma}\hat{N_\tau}| \Psi \rangle &=& N_\sigma N_\tau,
\end{eqnarray}
where $N_\sigma$ and $N_\tau$ are particle-number operators for particles of spin $\sigma$ and $\tau$, respectively, {\em i.e.},
\begin{equation}
    \hat{N}_\sigma = \sum_{p} \hat{a}^\dagger_{p_\sigma}\hat{a}_{p_\sigma}.
\end{equation}
These particle-number operators can also be used to generate constraints that connect RDMs of different ranks. For example, one can project $\hat{N_\tau}| \Psi \rangle = N_\tau | \Psi \rangle$ onto all subspaces defined by $\langle \Psi | \hat{a}^\dagger_{p_\sigma}\hat{a}_{q_\sigma}$ as 
\begin{align}
    \label{EQN:CONTRACT_D2_1}
         \forall p, q: \langle \Psi |\hat{a}_{p_\sigma}^\dagger \hat{a}_{q_\sigma}\hat{N_\tau}| \Psi \rangle =  N_\tau \langle \Psi |\hat{a}_{p_\sigma}^\dagger \hat{a}_{q_\sigma}| \Psi \rangle.
\end{align}
Equation \ref{EQN:CONTRACT_D2_1} defines contraction relationships between the spin-blocks of the 2RDM and the 1RDM. Lastly, high-spin eigenstates of $\hat{S}^2$ and $\hat{S}_z$ should satisfy, \cite{Mazziotti05_052505,Ayers12_014110} 
\begin{align}                                                                             \label{EQN:S2}
        \langle \Psi |\hat{S}^2| \Psi \rangle &= S(S+1).
\end{align}      
and 
\begin{equation}
    \label{EQN:MAX_SZ}
     \forall p, q: \langle \Psi | \hat{a}_{p_\sigma}^\dagger \hat{a}_{q_\tau} \hat{S}^+ | \Psi \rangle = 0.
\end{equation}
{\color{black}The constraints represented by Eqs.~\ref{EQN:TRACE1}, \ref{EQN:TRACE2}, \ref{EQN:CONTRACT_D2_1}, \ref{EQN:S2}, and \ref{EQN:MAX_SZ} can all be expressed in terms of the 1RDM and 2RDM by inserting the second-quantized definitions of the number and spin operators and bringing the resulting expressions to normal order with respect to the true vacuum.}

In addition to the straightforward equality constraints detailed above, more complex constraints govern allowable eigenvalues of various RDMs. Such constraints can be derived\cite{Erdahl01_042113} by defining a set of operators {$\hat{C}_I$} that generate basis functions from the wave function as
\begin{equation}                                                                             \label{EQN:BASIS}
        |\Psi_I \rangle = \hat{C}_I  |\Psi \rangle.
\end{equation}    
The {\color{black}Gram} matrix, ${\bf M}$, associated with this basis has elements
\begin{equation}                                                                                             \label{EQN:ANY_RDM}
        M^I_J =\langle \Psi_I|\Psi_J \rangle 
        =\langle \Psi| \hat{C}_I^{\dagger} \hat{C}_J  |\Psi \rangle
\end{equation}
and must be positive semidefinite. 
When {$\hat{C}_I$} is taken to be a single annihilation operator $\hat{a}_{p}$, it can be seen that the {\color{black}resulting Gram} matrix is the 1RDM. When {$\hat{C}_I$} is a pair of annihilation operators $\hat{a}_{p} \hat{a}_{q}$, the 2RDM is obtained. When {$\hat{C}_I$} is a creation operator $\hat{a}^{\dagger}_{p}$, a pair of creation operators $\hat{a}_{p} ^{\dagger} \hat{a}_{q} ^{\dagger}$, or a pair of creation-annihilation operators $\hat{a}_{p} ^{\dagger} \hat{a}_{q}$, one obtains the 1-hole RDM $^1{\bf Q}$, 2-hole RDM ${}^2{\bf Q}$, and particle-hole RDM ${}^2{\bf G}$ respectively. The positive semidefiniteness of ${}^2{\bf D}$, ${}^2{\bf Q}$, and ${}^2{\bf G}$ (plus ${}^1{\bf D}$ and ${}^1{\bf Q}$) constitutes the PQG conditions derived by Garrod and Percus.\cite{Percus64_1756}

\subsection{Three-particle $N$-representability conditions}
When {$\hat{C}_I$} takes the form of a three body operator, four additional unique RDMs can be defined:
\begin{equation}                                                                                             \label{EQN:D3}
    ^3D^{p_\sigma q_\tau r_\kappa}_{s_\sigma t_\tau u_\kappa}=
              \langle \Psi |
              \hat{a}^\dagger_{p \sigma}  \hat{a}^\dagger_{q \tau} \hat{a}_{r \kappa}^{\dagger}
              \hat{a}_{u \kappa}  \hat{a}_{t \tau}  \hat{a}_{s \sigma} 
              | \Psi \rangle,
\end{equation}
\begin{equation}                                                                                             \label{EQN:E3}
    ^3E^{p_\sigma q_\tau r_\kappa}_{ s_\lambda t_\mu u_\nu} =
              \langle \Psi |
              \hat{a}^\dagger_{p \sigma}  \hat{a}^\dagger_{q \tau} \hat{a}_{r \kappa}
              \hat{a}_{u \nu}^{\dagger}  \hat{a}_{t \mu}  \hat{a}_{s \lambda} 
              | \Psi \rangle,
\end{equation}
\begin{equation}                                                                                             \label{EQN:F3}
    ^3F^{p_\sigma q_\tau r_\kappa}_{ s_\lambda t_\mu u_\nu} =
            \langle \Psi |
            \hat{a}_{u \nu}^{\dagger}  \hat{a}_{t \mu}  \hat{a}_{s \lambda}
            \hat{a}^\dagger_{p \sigma}  \hat{a}^\dagger_{q \tau} \hat{a}_{r \kappa}
            | \Psi \rangle,
\end{equation}
and
\begin{equation}                                                                                             \label{EQN:Q3}
      ^3Q^{p_\sigma q_\tau r_\kappa}_{ s_\sigma t_\tau u_\kappa}=
              \langle \Psi |
              \hat{a}_{p \sigma}  \hat{a}_{q \tau} \hat{a}_{r \kappa}
              \hat{a}^\dagger_{u \kappa}  \hat{a}^\dagger_{t \tau}  \hat{a}^{\dagger}_{s \sigma} 
              | \Psi \rangle.
\end{equation}
{\color{black}Here, like $\sigma$ and $\tau$, the labels $\kappa$, $\lambda$, $\mu$, and $\nu$ indicate to $\alpha$- or $\beta$-spin functions.} For non-relativistic Hamiltonians, the non-zero spin blocks of ${}^3\mathbf{E}$ and ${}^3\mathbf{F}$ are those for which the number of $\alpha$-spin ($\beta$-spin) creation operators equals the number of $\alpha$-spin ($\beta$-spin) annihilation operators. Full ``3-positivity'' (or 3POS) requires the positive semidefiniteness of all four of these RDMs, as well as appropriate relations linking them to the 2RDM. 
{\color{black}Similar to what was done in Eq.~\ref{EQN:CONTRACT_D2_1} we can project $\hat{N_\kappa}| \Psi \rangle = N_\kappa | \Psi \rangle$ onto all subspaces defined by $\langle \Psi |\hat{a}_{p \sigma}^\dagger \hat{a}_{q \tau}^\dagger \hat{a}_{s \tau} \hat{a}_{r \sigma}$ to obtain}
\begin{equation}
    \label{EQN:CONTRACT_D3}
         \langle \Psi |\hat{a}_{p \sigma}^\dagger \hat{a}_{q \tau}^\dagger \hat{a}_{s \tau} \hat{a}_{r \sigma} \hat{N}_\kappa| \Psi \rangle =  N_\kappa \langle \Psi |\hat{a}_{p \sigma}^\dagger \hat{a}_{q \tau}^\dagger \hat{a}_{s \tau} \hat{a}_{r \sigma}| \Psi \rangle.
\end{equation}
{\color{black} The right-hand side of Eq.~\ref{EQN:CONTRACT_D3} is an element of the 2RDM (Eq.~\ref{EQN:D2}), scaled by $N_\kappa$. Inserting the second-quantized definition of $\hat{N}_\kappa$ into the left-hand side of Eq.~\ref{EQN:CONTRACT_D3} and bringing the resulting expression to normal order (relative to the true vacuum) then yields contraction relationships between the spin-blocks of the 3RDM (${}^3{\bf D}$) and the 2RDM.}

Due to the large floating-point and memory costs associated with manipulating the 3-body RDMs, 3POS constraints are rarely enforced. Because the sum of two positive semidefinite matrices are also positive semidefinite, weaker constraints based on the non-negativity of
\begin{align}
    \label{EQN:T1}
    \mathbf{T1} = {}^3\mathbf{D} + {}^3\mathbf{Q}\\
    \label{EQN:T2}
    \mathbf{T2} = {}^3\mathbf{E} + {}^3\mathbf{F}
\end{align}
can be defined,\cite{Erdahl78_697, Percus04_2095} with the T2 constraint being the stronger constraint of the two. One nice feature of these partial three-particle conditions is that the right-hand sides of Eqs.~\ref{EQN:T1} and \ref{EQN:T2} do not explicitly depend on any three-body RDMs.

\subsection{Electron correlation metrics}
\label{SEC:metrics}

A variety of metrics have been put forth to quantify the degree of electron correlation in many electron systems; these metrics are often defined in terms of one- or two-body quantities. 
In this work, at the one-particle level, we consider the von Neumann entropy, which borrows concepts from information theory, \cite{Milburn08_860}
quantifying the degree of correlation as
\begin{equation}                \label{EQN:ENTROPY}
    S({}^1{\bf D}) = -\sum_{p}{n_p {\rm ln}(n_p)}.
\end{equation}
Here, $n_p$ is the occupation of the $p$th natural spin-orbital, and the natural spin orbitals are obtained as the eigenfunctions of ${}^1{\bf D}$. The von Neumann entropy is zero for an uncorrelated wave function, where all $n_p$ are zero or one; it exhibits its maximum value when all spin-orbitals are partially occupied, with equal occupations. 

At the two-particle level, correlation metrics are generally expressed in terms of the two-particle cumulant RDM (or 2-cumulant, ${}^2{ \lambda}$), which is the portion of the 2RDM that is not expressible in terms of the 1RDM:\cite{Mukherjee97_432, Mazziotti98_419, Harriman02_7464}
\begin{equation}                             \label{EQN:Del2}
    {}^2\lambda^{p_\sigma q_\tau}_{r_\sigma s_\tau} = {}^2D^{p_\sigma q_\tau}_{r_\sigma s_\tau} - {}^1D^{p_\sigma}_{r_\sigma} {}^1D^{q_\tau}_{s_\tau} + \delta_{\sigma \tau}{}^1D^{p_\sigma}_{s_\tau} {}^1D^{q_\tau}_{r_\sigma}
\end{equation}
The elements of the 2-cumulant associated with an uncorrelated wave function are all zero, and thus the magnitude of this matrix, as measured by the
Frobenius norm
\begin{equation}                                 \label{EQN:2Norm}
    ||{}^2{ \lambda}|| = \sqrt{\sum_{\sigma \tau}\sum_{pqrs}|{}^2\lambda^{p_\sigma q_\tau}_{r_\sigma s_\tau}|^2}
\end{equation}
can quantify the degree of correlation in the system.\cite{Prezhdo05_582, Torre10_144104,Kais06_2543, Mazziotti06_174105, Prezhdo07_2879} The 2-cumulant has the following nice properties that enhance its utility as a correlation metric: (i) its Frobenius norm is invariant to unitary transformations and (ii)  the trace and square of the Frobenius norm of the 2-cumulant are additive.\cite{Mazziotti06_174105, Torre10_144104,Schaefer20_6150}

To our knowledge, no studies have considered the information contained in the three-body cumulant RDM (the 3-cumulant, ${}^3{ \lambda}$) as a measure of correlation in many-body systems. This quantity contains information regarding pure three-body correlation effects, the magnitude of which can be quantified via the Frobenius norm of the 3-cumulant
\begin{equation}
\label{EQN:3Norm}
||{}^3{ \lambda}|| = \sqrt{\sum_{\sigma \tau \kappa} \sum_{pqrstu}|{}^2\lambda^{p_\sigma q_\tau r_\kappa}_{s_\sigma t_\tau u_\kappa}|^2}
\end{equation}
where\cite{Mukherjee99_2800, Mazziotti98_419, Harriman02_7464}
\begin{align}                                                                                             \label{EQN:Del3}
    {}^3\lambda^{p_\sigma q_\tau r_\kappa}_{s_\sigma t_\tau u_\kappa} & =  {}^3D^{p_\sigma q_\tau r_\kappa}_{s_\sigma t_\tau u_\kappa} \nonumber \\
                          & - {}^1D^{p_\sigma}_{s_\sigma}{}^2\lambda^{q_\tau r_\kappa}_{t_\tau u_\kappa} 
                          +\delta_{\sigma \tau} {}^1D^{p_\sigma}_{t_\tau}{}^2\lambda^{q_\tau r_\kappa}_{s_\sigma u_\kappa} 
                          +\delta_{\sigma \kappa} {}^1D^{p_\sigma}_{u_\kappa}{}^2\lambda^{q_\tau r_\kappa}_{t_\tau s_\sigma} \nonumber \\
                          & - {}^1D^{q_\tau}_{t_\tau}{}^2\lambda^{p_\sigma r_\kappa}_{s_\sigma u_\kappa} 
                          +\delta_{\sigma \tau} {}^1D^{q_\tau}_{s_\sigma}{}^2\lambda^{p_\sigma r_\kappa}_{t_\tau u_\kappa} 
                          +\delta_{\tau \kappa} {}^1D^{q_\tau}_{u_\kappa}{}^2\lambda^{p_\sigma r_\kappa}_{s_\sigma t_\tau} \nonumber \\
                          & - {}^1D^{r_\kappa}_{u_\kappa}{}^2\lambda^{p_\sigma q_\tau}_{s_\sigma t_\tau} 
                          +\delta_{\sigma \kappa} {}^1D^{r_\kappa}_{s_\sigma}{}^2\lambda^{p_\sigma q_\tau}_{u_\kappa t_\tau} 
                          +\delta_{\tau \kappa} {}^1D^{r_\kappa}_{t_\tau}{}^2\lambda^{p_\sigma q_\tau}_{s_\sigma u_\kappa} \nonumber \\
                          & - {}^1D^{p_\sigma}_{s_\sigma}{}^1D^{q_\tau}_{t_\tau}{}^1D^{r_\kappa}_{u_\kappa} \nonumber \\
                          & -\delta_{\sigma \tau}\delta_{\sigma \kappa} {}^1D^{p_\sigma}_{t_\tau}{}^1D^{q_\tau}_{u_\kappa}{}^1D^{r_\kappa}_{s_\sigma}\nonumber \\
                          & -\delta_{\sigma \tau}\delta_{\sigma \kappa} {}^1D^{p_\sigma}_{u_\kappa}{}^1D^{q_\tau}_{s_\sigma}{}^1D^{r_\kappa}_{t_\tau} \nonumber \\
                          & +\delta_{\tau \kappa} {}^1D^{p_\sigma}_{s_\sigma}{}^1D^{q_\tau}_{u_\kappa}{}^1D^{r_\kappa}_{t_\tau}\nonumber \\
                          & +\delta_{\sigma \kappa} {}^1D^{p_\sigma}_{u_\kappa}{}^1D^{q_\tau}_{t_\tau}{}^1D^{r_\kappa}_{s_\sigma}\nonumber \\
                          & +\delta_{\sigma \tau} {}^1D^{p_\sigma}_{t_\tau}{}^1D^{q_\tau}_{s_\sigma}{}^1D^{r_\kappa}_{u_\kappa}                          
\end{align}
As in the case of the 2-cumulant, the Frobenius norm of the 3-cumulant is invariant to unitary transformations, and ${\rm Tr}({}^3{ \lambda})$ and $||{}^3{ \lambda}||^2$ are additive.

\subsection{Spin correlation metrics}

We also consider the possibility of spin frustration, or the inability to maximize antiferromagnetic interactions, in the H$_{10}$ model systems. We quantify the manifestation of spin frustration using a single scalar quantity: the sum of the absolute values of site-wise spin correlations\cite{Evangelista20_104018,Scuseria14_9925}
\begin{equation}
\label{EQN:S2ABS}
    \langle \hat{S}^2 \rangle_{\rm abs} = \sum_{ij} | \langle \mathbf{\hat{S}}_i \cdot \mathbf{\hat{S}}_j \rangle |.
\end{equation}
Here, $\mathbf{\hat{S}}_i$ represents the spin operator associated with localized orbital, $\phi_i$, which is localized using the Pipek-Mezey procedure.\cite{Mezey89_4916} This quantity is expressible in terms of the 1RDM and 2RDM and reduces to the usual spin-squared expectation value, $\langle \hat{S}^2 \rangle$, if the absolute value in Eq.~\ref{EQN:S2ABS} is lifted.

\section{Computational Details}

\label{SEC:COMPUTATIONAL_DETAILS}

Following Ref.~\citenum{Evangelista20_104018},  electronic energies of 1-dimensional (chain, ring), 2-dimensional (sheet), and 3-dimensional (pyramid)  H$_{10}$ clusters were evaluated at the full CI and v2RDM levels of theory. All geometries were obtained from the  GitHub repository associated with Ref.~\citenum{Evangelista20_104018} (see Ref.~\citenum{Evangelista20_github}). All full CI calculations were performed using the GAMESS package, \cite{Gordon20_154102} and full CI RDMs were computed using an in-house code from wave function expansion coefficients generated by GAMESS.  All v2RDM optimizations were carried out using \texttt{hilbert},\cite{hilbert} which is a plugin to the \textsc{Psi4} package.\cite{Sherrill20_184108} A similar implementation can be found in the version 5.4 release of the Q-Chem package.\cite{HeadGordon15_184}. All full CI and v2RDM calculations were carried out within the STO-6G basis set,\cite{Pople69_2657} within the basis of restricted Hartree--Fock molecular orbitals. 

Additional calculations on small-molecule systems were performed at the CI- and v2RDM-driven CASSCF levels of theory.  We refer the reader to Refs.~\citenum{Mazziotti08_134108} and \citenum{DePrince16_2260} for a description of how the v2RDM approach can be applied to active-space-based computations. All CI- and v2RDM-CASSCF calculations used full-valence active spaces, and the electron-repulsion integrals were represented using the density fitting approximation;\cite{Whitten73_4496,Sabin79_3396} the primary and auxiliary basis sets were cc-pVQZ\cite{Dunning89_1007} and cc-pVQZ-jk,\cite{Weigend02_4285} respectively. CI-based CASSCF calculations were performed using the \textsc{Psi4} package, and v2RDM-based CASSCF calculations were carried out using \texttt{hilbert}.

\section{Results and Discussion}

\label{SEC:RESULTS}

 \begin{figure*}[!htpb]
    \caption{Structures of the H$_{10}$ clusters considered in this work; each structure is characterized by the nearest-neighbor distance, $r$.  }
    \label{FIG:STRUCTURES}
    \begin{center}
        \includegraphics[width=6.0in]{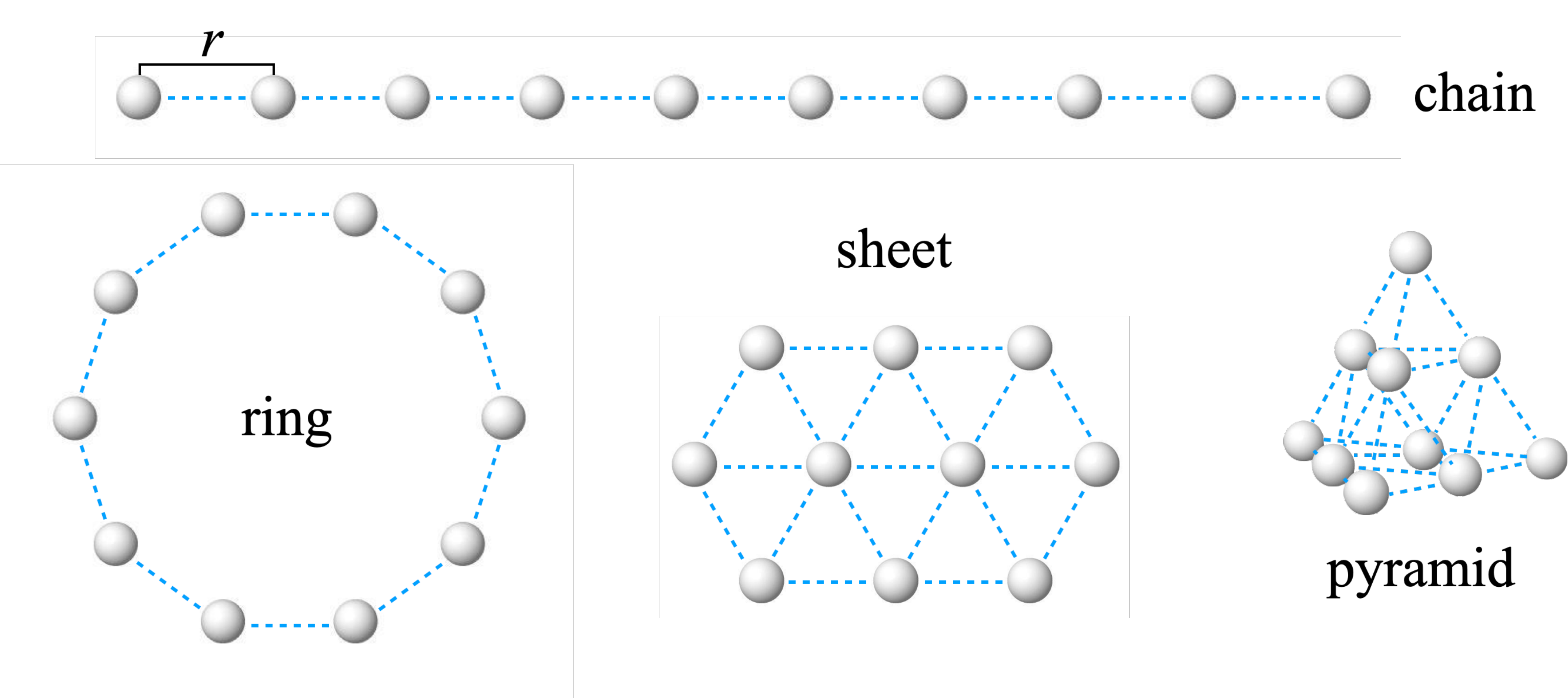}
    \end{center}
\end{figure*}

\begin{figure}[h]

    \caption{Errors in v2RDM energies (mE$_{\rm h}$) relative to those from full CI for the (a) chain, (b) ring, (c) sheet, and (d) pyramid H$_{10}$ clusters.}
    \label{FIG:ENERGY}
    \begin{center}
        \includegraphics[scale=1.0]{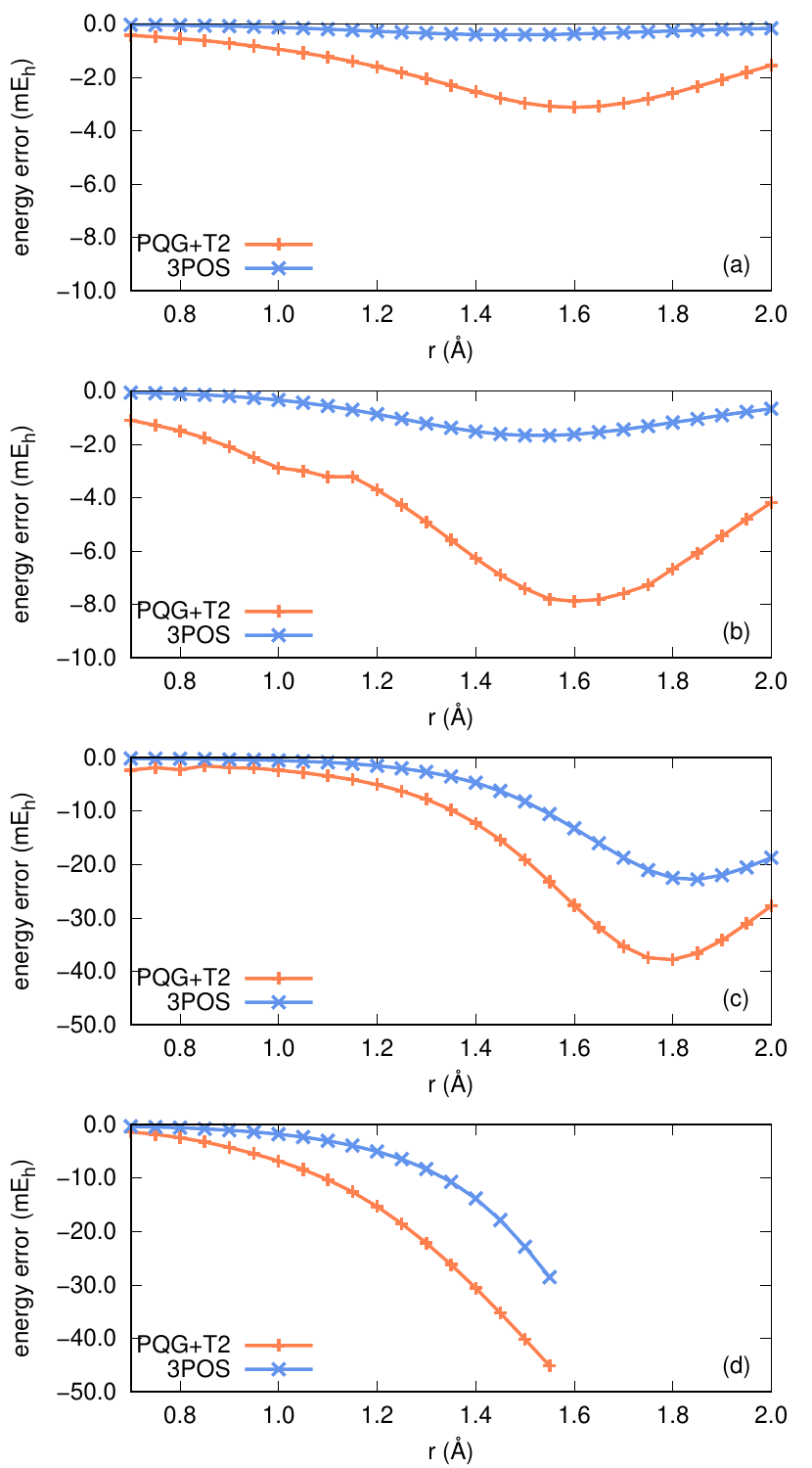}
    \end{center}

\end{figure}

We now assess the accuracy of v2RDM-derived energies for 1-, 2-, and 3-dimensional H$_{10}$ clusters when optimized RDMs satisfy partial or full three-particle $N$-representability conditions.  The geometries of these structures are depicted in Fig.~\ref{FIG:STRUCTURES}; for a more detailed description of these geometries, we refer the reader to Refs.~\citenum{Evangelista20_104018} and \citenum{Evangelista20_github}. Figure \ref{FIG:ENERGY} illustrates errors in energies obtained from v2RDM calculations performed under PQG+T2 and 3POS conditions, with energies from full CI serving as reference values. Here, the label ``$r$'' refers to nearest-neighbor H--H distance.  In the case of the 3-dimensional (pyramid) structure, we only consider H--H distances up to 1.55 \AA~because, beyond this distance, a state-crossing occurs at the full CI level. With the conditions employed in this work, the v2RDM approach can only describe the ground state of a given spin symmetry; comparing v2RDM results to energies for two different CI states complicates our analysis. 

Inspection of Fig.~\ref{FIG:ENERGY} reveals the following. First, because PQG+T2 and 3POS conditions are necessary yet insufficient conditions for ensemble $N$-representability of the 2RDM, v2RDM provides a lower bound to the full CI energy at all geometries, and energies from 3POS calculations always provide the better lower-bound. Second, for both of the 1-dimensional models (H$_{10}$ chain and ring), 3POS calculations are highly accurate, exhibiting errors no larger than 1.7 mE$_{\rm h}$. PQG+T2 calculations are slightly less accurate for these 1-dimensional systems, with a maximum error of nearly 8 mE$_{\rm h}$ in the case of the ring geometry. On the other hand, v2RDM yields significantly less accurate energy estimates for both the sheet and pyramid geometries. This behavior was reported previously in Ref.~\citenum{Evangelista20_104018} at the PQG+T2 level of theory. Here, we note that 3POS energies are somewhat more accurate than those from PQG+T2, but errors exceeding 20 mE$_{\rm h}$ are still observed for both the sheet and pyramid structures.  

\begin{table}[]
\centering
\caption{Non-parallelity errors (E$_{\rm h}$) for v2RDM-based descriptions of potential energy curves for 1-, 2-, and 3-dimensional H$_{10}$ clusters.}
\label{TABLE:NONPARALLELITY_ERROR}
\begin{tabular}{lccccc}
\hline
\hline
        & PQG     & PQG+T2   & 3POS   \\
      \hline 
chain   & 0.0228 & 0.0027  & 0.0004 \\
ring    & 0.0382 & 0.0074  & 0.0016 \\
sheet   & 0.1108 & 0.0363  & 0.0226 \\
pyramid & 0.1073 & 0.0438  & 0.0283 \\
\hline
\hline
\end{tabular}
\end{table}

Table \ref{TABLE:NONPARALLELITY_ERROR} provides non-parallelity errors (NPEs) for v2RDM-derived energies, where the NPE is defined as the difference between the maximum and minimum absolute deviations of v2RDM energies from full CI ones. The NPE are evaluated for the range 0.70 \AA--2.00 \AA~for the ring, chain, and sheet geometries and 0.70 \AA--1.55\AA~for the pyramid geometry. Again, we find that 3POS is highly accurate for the chain and ring models (with NPE values of only 0.4 and 1.6 mE$_{\rm h}$, respectively), while larger NPE values are observed for the sheet and pyramid structures (22.6 and 28.3 mE$_{\rm h}$, respectively). PQG+T2 constraints yield somewhat larger NPE values for all H$_{10}$ models, while two-particle constraints alone (PQG) lead to significantly less accurate potential energy curves. The NPE values for v2RDM with PQG conditions exceed 100 mE$_{\rm h}$ for both the sheet and pyramid structures.

\begin{figure*}[!htpb]
    \caption{Various correlation metrics, including (a) the correlation energy, (b) the non Neumann entropy, (c) the square magnitude two-particle cumulant RDM, (d) the square magnitude of the three-particle cumulant RDM, and (e) the ratio  $||{}^3\lambda||^2 / ||{}^2\lambda||^2$ {\color{black} computed at the full CI level of theory. Panels (f)-(j) depict the same metrics, evaluated at the v2RDM/3POS level of theory.} }
    \label{FIG:CUMULANT}
    \begin{center}
        \includegraphics[scale=1.0]{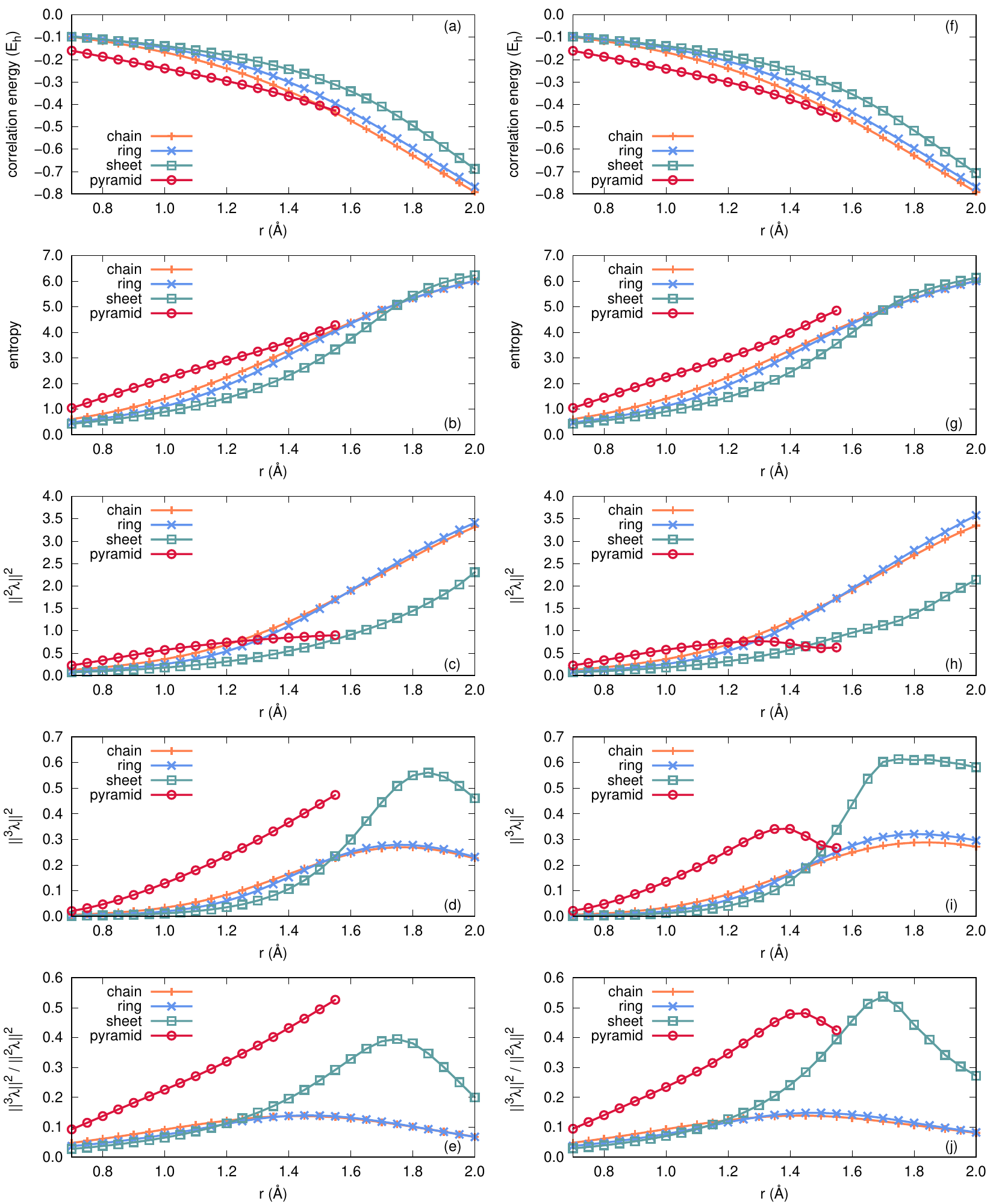}
    \end{center}

\end{figure*}

Having established that 3POS improves upon PQG+T2 but nevertheless has large errors associated with it in certain cases, it is desirable to develop a correlation metric that could indicate the reliability of energies from v2RDM calculations performed under 3POS conditions. Figure \ref{FIG:CUMULANT} depicts several possible {\color{black}full-CI-based} correlation metrics {\color{black}that depend upon} one-, two-, or three-particle quantities. Panel (a) of Fig.~\ref{FIG:CUMULANT} illustrates the correlation energy (defined as the difference between the {\color{black}full CI} and restricted Hartree-Fock energies) for each system as a function of nearest-neighbor distance, $r$. For each system, the magnitude of the correlation energy increases with increasing $r$, and the largest-magnitude correlation energies are associated with the chain and ring geometries at 2.0 \AA. Neither of these properties correlate well with the energy error exhibited by 3POS; the errors do not increase monotonically, and the largest errors occur in the spin recoupling region, well before $r = 2.0$ \AA. Moreover, 3POS errors for the sheet geometry are significantly larger than those for the chain and ring structures, despite the correlation energy suggesting that the latter are the most correlated systems. Panel (b) of Fig.~\ref{FIG:CUMULANT} depicts the {\color{black}full CI} von Neumann entropy, which also correlates poorly with the energy errors associated with 3POS. As with the correlation energy, the entropy increases monotonically with increasing $r$, and the chain, ring, and sheet geometries all appear to be similarly correlated at $r=2.0$ \AA, according to this metric. The squared magnitude of the {\color{black}full CI} 2-cumulant is depicted in panel (c) of Fig.~\ref{FIG:CUMULANT}. Again, this metric does not correlate well with the 3POS energy errors. $||{}^2\lambda||^2$ increases monotonically with $r$, and the chain and ring structures appear to be far more correlated than the sheet structure at large $r$, according to this metric. 

The squared magnitudes of the 3-cumulant presented in panel (d) display fundamentally different behaviors. First, it is clear that the $||{}^3\lambda||^2$ values, like the 3POS energy errors, do not increase monotonically with $r$; rather, they exhibit maxima at stretched geometries. However,  the $r$ values at which these maxima occur do not coincide with the $r$ values at which we observe the largest energy errors from 3POS{\color{black}, particularly for the chain and ring structures}. Second, $||{}^3\lambda||^2$ values for the pyramid at small and intermediate $r$ are significantly larger than those for other structures at similar $r$ values, which is consistent with the relative energy errors for the pyramid geometry, as compared to the other structures. Third, $||{}^3\lambda||^2$ values for the sheet geometry at large $r$ are significantly larger than those associated with the chain and ring structures, which is also consistent with the behavior of the 3POS energy errors in Fig.~\ref{FIG:ENERGY}. These observations lead us to suspect that larger energy errors in 3POS could be linked to large-magnitude three-body correlations. In panel (e) we illustrate the ratio of $||{}^3\lambda||^2$ to $||{}^2\lambda||^2$, which we interpret as the relative importance of pure three-body and two-body correlations. Here, we observe some similarities to the behavior of $||{}^3\lambda||^2$ depicted in panel (d), with the primary difference being that the maxima in the curves associated with the chain and ring geometries are shifted to smaller $r$. For the chain, ring, and sheet geometries, the maxima in $||{}^3\lambda||^2 / ||{}^2\lambda||^2$ correlate reasonably well with the largest-magnitude energy errors for 3POS depicted in Fig.~\ref{FIG:ENERGY}. 

{\color{black} Given the prohibitive cost of full CI calculations, correlation metrics based on this approach will not be useful indicators for the reliability of approximate methods such as v2RDM in general systems. As such, we evaluated the v2RDM/3POS analogues of the metrics depicted in panels (a)-(e) of Fig.~\ref{FIG:CUMULANT} and depict them in panels (f)-(j).  The same general conclusions can be drawn regarding the poor correlation between the energy error associated with 3POS and the correlation energy [panel (f)], the von Neumann entropy [panel (g)], the square magnitude of the two-cumulant RDM [panel (h)], and the magnitude of the two-cumulant RDM [panel (i)]. Moreover, as can be seen in panel (j), the $r$ values at which the maximal $||{}^3\lambda||^2 / ||{}^2\lambda||^2$ values occur agree well with those depicted in panel (e). The only exception is the pyramid system for which v2RDM/3POS predicts a maximum in $||{}^3\lambda||^2 / ||{}^2\lambda||^2$, whereas full CI predicts that this quantity increase monotonically, at least up to 1.55 \AA. From these data, we conclude the following: (i) of all correlation metrics considered, $||{}^3\lambda||^2 / ||{}^2\lambda||^2$ appears to correlate the best with energy errors associated with 3POS calculations, and (ii) the 3POS values of $||{}^3\lambda||^2 / ||{}^2\lambda||^2$ are similar enough to those obtained from full CI that the former can serve as a useful proxy for the latter. Hence, for the remainder of this work, we focus on the metric $||{}^3\lambda||^2 / ||{}^2\lambda||^2$ computed at the v2RDM/3POS level of theory. }

\begin{figure*}[!htpb]
    \caption{(a) Absolute energy errors (mE$_{\rm h}$) for v2RDM calculations under 3POS conditions as a function of $||{}^3\lambda||^2 / ||{}^2\lambda||^2$ and (b) a zoomed-in perspective of the data in the box indicated by dashed lines in panel (a); (c) percent errors in the v2RDM correlation energy and (d) a zoomed-in perspective of the area indicated by dashed lines in panel (c).  }
    \label{FIG:CORRELATION_PLOT}
    \begin{center}
        \includegraphics{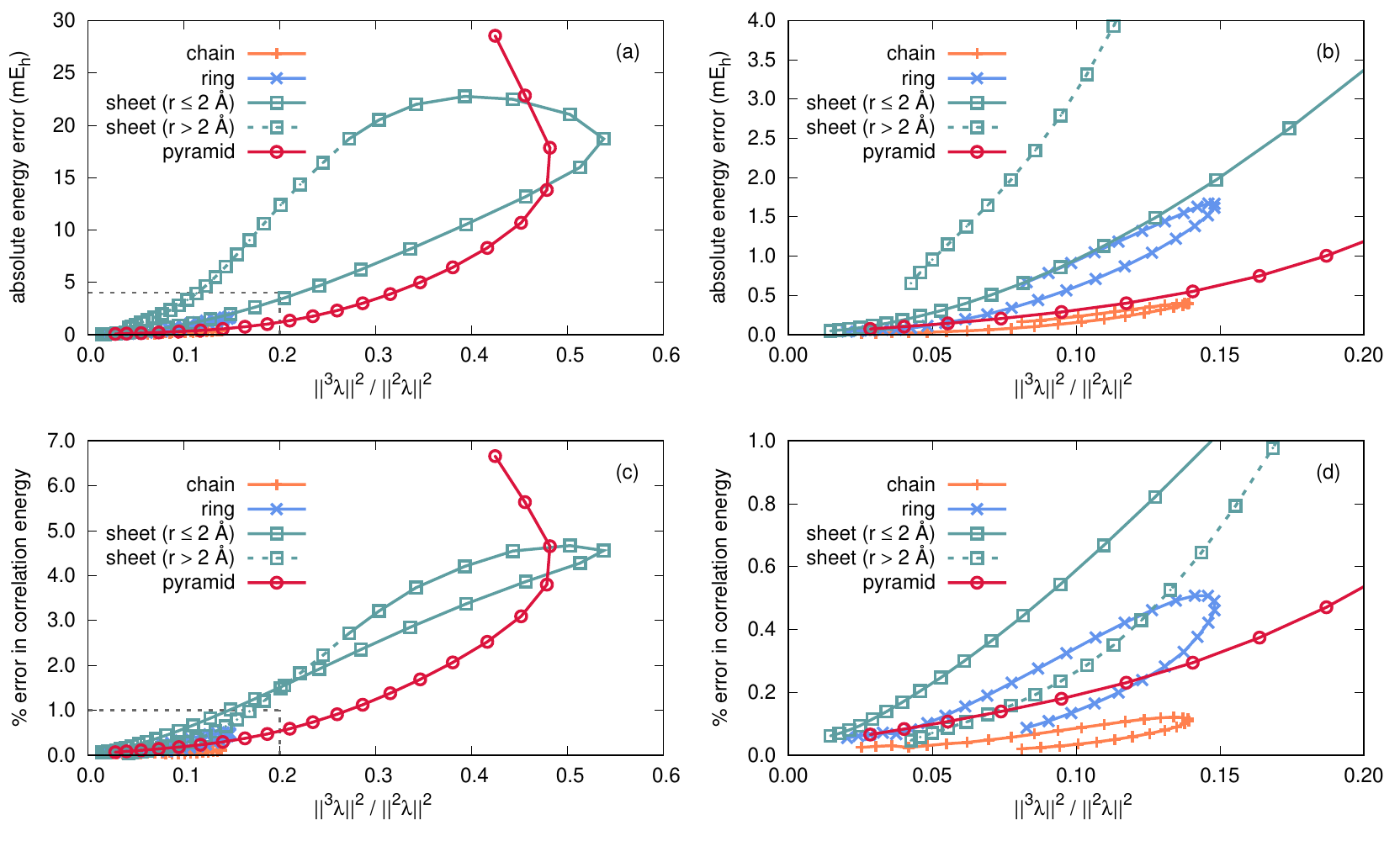}
    \end{center}
\end{figure*}

Panels (a) and (b) of Fig.~\ref{FIG:CORRELATION_PLOT} plot the absolute energy error from v2RDM calculations with 3POS conditions against the ratio $||{}^3\lambda||^2 / ||{}^2\lambda||^2$ {\color{black}computed at the same level of theory}. Note that additional data for the sheet structure are included in this figure and the following analysis, in the range 2 \AA~< $r$ $\le$ 3 \AA. From panel (a), we can see that the statistical correlation between these two quantities is poor, at least for the sheet and pyramid geometries (simple linear fits to these data result in R$^2$ values of 0.74 and 0.63, respectively).  Panel (b) of Fig.~\ref{FIG:CORRELATION_PLOT} provides a more detailed perspective of the region indicated by the dashed box in panel (a). Here, we find that, for the chain and ring geometries, the correlation between energy error and the squared cumulant norm ratio is much better; linear fits to these data result in R$^2$ values of 0.93 and 0.94 respectively. However, the rate of change of the energy error with respect to $||{}^3\lambda||^2 / ||{}^2\lambda||^2$ is not consistent across structures, which, along with the poor R$^2$ values calculated for the sheet and pyramid structures, suggests that this the squared cumulant norm ratio would not be suitable as a universal metric for quantifying energy error in 3POS calculations. Indeed, the R$^2$ value resulting from a linear fit to the entire data set is only 0.69. Nonetheless, we can conclude that, for these systems, large values of $||{}^3\lambda||^2 / ||{}^2\lambda||^2$ are associated with large energy errors. For the sheet and pyramid structures, geometries at which $||{}^3\lambda||^2 / ||{}^2\lambda||^2$ exceeds 0.3 have absolute energy errors associated with them that are larger than 2 kcal/mol and 5 kcal/mol, respectively. However, the converse cannot be stated; a value of $||{}^3\lambda||^2 / ||{}^2\lambda||^2 < 0.3$ does not guarantee a small energy error. 

Panels (c) and (d) of Fig.~\ref{FIG:CORRELATION_PLOT} depict a slightly different representation of the energy error, the percent error in the correlation energy captured by v2RDM theory, as a function of the ratio $||{}^3\lambda||^2 / ||{}^2\lambda||^2$. Here, the correlation energy is defined as the difference between the full CI and restricted Hartree-Fock energies, within the STO-6G basis set. We find that the correlation between these two quantities is, overall, slightly better than in the case of the data presented in panel (a). Here, a simple linear fit to the total data set results in an R$^2$ value of 0.85; this improved correlation relative to that for the absolute energy error versus $||{}^3\lambda||^2 / ||{}^2\lambda||^2$ is due to the excellent correlation between the percent error and $||{}^3\lambda||^2 / ||{}^2\lambda||^2$ for the sheet structures (R$^2$=0.97). On the other hand the R$^2$ values for the data corresponding to the chain and ring structures [see panel (d)] are less good than described above (0.59 and 0.71, respectively). The R$^2$ value for the data corresponding to the pyramid structure are similar using either measure of the energy error (in this case 0.72). As above, values of $||{}^3\lambda||^2 / ||{}^2\lambda||^2$ > 0.3 are associated with larger errors. For the sheet and pyramid structures, geometries at which $||{}^3\lambda||^2 / ||{}^2\lambda||^2$ exceeds 0.3 are associated with percent correlation energy errors  that are larger than 2.8\% and 1.3\%, respectively.

Before moving on, we briefly consider the manifestation of spin frustration in the H$_{10}$ model systems, which was studied in detail in Ref.~\citenum{Evangelista20_104018}. To summarize those findings, clear antiferromagetic ordering is observed for the chain and ring structures, while the geometries of the sheet and pyramid structures preclude such long-range ordering. Decreased spin-spin correlations in the 2- and 3-dimensional structures can be quantified in several ways, including the sum of the absolute value of the site-wise spin-spin correlations. Table \ref{TAB:SPIN_FRUSTRATION} tabulates $\langle \hat{S}^2\rangle_{\rm abs}$ (see Eq.~\ref{EQN:S2ABS}) at the CI and v2RDM levels of theory, calculated for the four H$_{10}$ model systems at a nearest neighbor H--H separation of 1.5 \AA. CI results clearly indicate a larger degree of spin-spin correlations in the chain and ring structures, relative to that observed for the sheet and pyramid structures. These trends are generally reproduced by v2RDM with two-particle $N$-representability conditions (PQG), although the quantitative agreement between PQG and CI results is poor. Better agreement with CI can be obtained using PQG+T2 constraints. Results obtained using full three-particle conditions (3POS) only improve slightly upon those from PQG+T2 calculations, and we observe that 3POS results in a clear systematic overestimation of the spin-spin correlation for all structures. The degree to which 3POS overestimates $\langle S^2 \rangle_{\rm abs}$ is slightly smaller for the chain, ring, and sheet structures (0.13, 0.10, and 0.19 respectively), as compared to the pyramid structure (0.51). Despite these quantitative differences, 3POS does a good job of clearly delineating frustrated and non-frustrated systems and also predicts the correct ordering of $\langle S^2 \rangle_{\rm abs}$ values overall.

\begin{table}[!htpb]
\centering
\caption{The sum of the absolute value of the site-wise spin-spin correlation, $\langle S^2 \rangle_{\rm abs}$, for different H$_{10}$ clusters at a nearest neighbor H--H distance of 1.5 \AA. }
\label{TAB:SPIN_FRUSTRATION}
\begin{tabular}{lcccc}
\hline
\hline
           &     CI &  PQG   & PQG+T2 &   3POS \\ 
\hline
chain      &  17.42 &  18.50 &  17.67 &  17.55 \\
ring       &  18.66 &  17.74 &  18.54 &  18.76 \\
sheet      &  11.55 &  12.39 &  11.91 &  11.74 \\
pyramid    &  10.86 &  12.12 &  11.41 &  11.37 \\
\hline
\hline
\end{tabular}
\end{table}

It remains to be seen whether the conclusions we have drawn regarding the relationship between $||{}^3\lambda||^2 / ||{}^2\lambda||^2$ and the energy errors of 3POS are translatable to other molecular systems. Table \ref{TAB:MOLECULES} tabulates energy errors from full-valence v2RDM-driven CASSCF calculations\cite{Mazziotti08_134108,DePrince16_2260} for several small molecules at their equilibrium geometries,\cite{Lide05_2005} within the cc-pVQZ basis set; errors are evaluated relative to energies from CI-based CASSCF calculations.  v2RDM theory provides a reasonable description of these systems when optimized 2RDMs satisfy the two-particle (PQG) conditions; in this case, energy errors are generally on the order of 10$^{-3}$--10$^{-2}$ E$_{\rm h}$, with the largest observed error being -13.5 mE$_{\rm h}$ for CH$_4$. This result, in particular, is consistent with the results of Ref.\citenum{Mazziotti06_032501}, which considered v2RDM- and full-CI based descriptions of these same molecules in a minimal basis set. The PQG error for CH$_4$ was reported to be -11.7 mE$_{\rm h}$ in that work. Unsurprisingly, partial three-particle conditions (T2) significantly reduce the energy errors associated with PQG calculations, with the largest observed error being only -0.7 mE$_{\rm h}$, in the case of CO. The application of full three-particle conditions results in nearly exact energies; the only molecule for which we observe an absolute energy error larger than 10$^{-4}$ E$_{\rm h}$ is H$_2$O (-0.3 mE$_{\rm h}$). This high accuracy is, again, consistent with the minimal-basis results of Ref.~\citenum{Mazziotti06_032501}. These results indicate that the high accuracy of 3POS previously observed in minimal-basis calculations is retained in active-space-based calculations such as v2RDM-driven CASSCF. Table \ref{TAB:MOLECULES} also includes the ratio $||{}^3\lambda||^2 / ||{}^2\lambda||^2$ determined from v2RDM-derived RDMs that satisfy the 3POS conditions. We find that this quantity is small for all systems considered here, well-below the threshold for concern (0.3) identified above for hydrogen clusters.

\begin{table}[]
\centering
\caption{CASSCF energies (E$_{\rm h}$), the deviation of v2RDM energies from these reference values (mE$_{\rm h}$), and values of $||{}^3\lambda||^2 / ||{}^2\lambda||^2$ determined from 3POS calculations. }
\label{TAB:MOLECULES}
\begin{tabular}{lccccc}
\hline
\hline
          &                   & \multicolumn{3}{c}{energy error} &  \\ 
          \cline{3-5}  
molecule    & CASSCF       & PQG      & PQG+T2    & 3POS         & $||{}^3\lambda||^2 / ||{}^2\lambda||^2$  \\ \hline
BeH$_2$     & -15.8142     & -0.5     & 0.0       & 0.0          & 0.022                                    \\
BH          & -25.1873     & -1.3     & 0.0       & 0.0          & 0.076                                    \\
CH$_4$      & -40.2994     & -13.5    & -0.5      & 0.0          & 0.022                                    \\
CO          & -112.921     & -8.4     & -0.7      & 0.0          & 0.047                                    \\
H$_2$O      & -76.1182     & -2.2     & -0.3      & -0.3         & 0.020                                    \\
NH$_3$      & -56.2972     & -5.3     & -0.5      & 0.0          & 0.021                                    \\

\hline
\hline
\end{tabular}
\end{table}

Lastly, we consider the potential energy curves for the dissociation of molecular nitrogen and carbon monoxide. Table \ref{TAB:N2} presents energy errors from v2RDM-based CASSCF calculations at several N--N and C--O distances. As above, all calculations used the cc-pVQZ basis set and a full-valence active space. Energy errors are with respect to energies obtained from CI-based CASSCF, and v2RDM-driven CASSCF calculations were carried out under 3POS conditions. Total energies at the CI- and v2RDM-CASSCF levels of theory can be found in the Supporting Information. We find that 3POS provides highly accurate energies at all geometries, with maximum errors of only 1.1 mE$_{\rm h}$ and 0.6 mE$_{\rm h}$ at an N--N distance of 1.7 \AA~and a C--O distance of 2.0 \AA, respectively.  For N$_2$, $||{}^3\lambda||^2 / ||{}^2\lambda||^2$ values evaluated using cumulant RDMs that satisfy 3POS conditions are modest at all geometries; the maximum observed value (0.105) occurs at the same N--N distance as the largest 3POS error (1.7 \AA). So, for non-equilibrium geometries in this particular molecule, our general conclusions from above hold: 3POS accurately reproduces CI-derived energies, and the relative magnitudes of three- and two-body correlations, as measured by the ratio of the squared norms of the relevant cumulant RDMs, correlates reasonably well with the energy error. On the other hand, $||{}^3\lambda||^2 / ||{}^2\lambda||^2$ values computed for CO are generally much larger and do not peak at the same C--O distance for which we observe the largest energy error (2.0 \AA). The maximum value of $||{}^3\lambda||^2 / ||{}^2\lambda||^2$ we observe is 0.427 at a C--O distance of 2.5 \AA; at this geometry, the 3POS energy error is only 0.2 mE$_{\rm h}$. Clearly, the metric $||{}^3\lambda||^2 / ||{}^2\lambda||^2$ does not correlate well with the energy error in this case.

\begin{table}[]
\centering
\caption{ Errors in v2RDM-driven CASSCF energies (mE$_{\rm h}$) relative to CI-based CASSCF energies for N$_2$ and CO at several interatomic distances (r, \AA), as well as v2RDM-derived $||{}^3\lambda||^2 / ||{}^2\lambda||^2$ values. RDMs from v2RDM-based calculations satisfy the 3POS conditions.}
\label{TAB:N2}
\begin{tabular}{lccccc}
\hline
\hline

           & \multicolumn{2}{c}{N$_2$}   &~ & \multicolumn{2}{c}{CO}    \\ 
           \cline{2-3} \cline{5-6}
$r$ & error & $||{}^3\lambda||^2 / ||{}^2\lambda||^2$ &~& error & $||{}^3\lambda||^2 / ||{}^2\lambda||^2$  \\ \hline
1.0        & 0.0  & 0.041  &~& 0.0  & 0.034  \\
1.1        & 0.0  & 0.052  &~& 0.0  & 0.044  \\
1.2        & 0.1  & 0.064  &~& 0.0  & 0.056  \\
1.3        & 0.2  & 0.078  &~& 0.1  & 0.071  \\
1.5        & 0.6  & 0.103  &~& 0.2  & 0.112  \\
1.7        & 1.1  & 0.105  &~& 0.4  & 0.167  \\
2.0        & 0.5  & 0.050  &~& 0.6  & 0.267  \\
2.5        & 0.0  & 0.010  &~& 0.2  & 0.427  \\

\hline
\hline
\end{tabular}
\end{table}

\section{Conclusions}

\label{SEC:CONCLUSIONS}

Variational two-electron reduced-density-matrix theory offers a formally polynomially-scaling and systematically-improvable description of the electronic structure of many-electron systems in which nondynamic correlation effects dominate. In particular, v2RDM calculations performed under partial\cite{Percus04_2095} and full\cite{Mazziotti06_032501} three-particle $N$-representability often closely reproduce full-CI results, at least for small molecules, at both equilibrium and non-equilibrium geometries.\cite{Mazziotti06_032501,DePrince16_2260} On the other hand, recent benchmark studies on hydrogen cluster models have revealed challenges for v2RDM carried out under partial three-particle (PQG+T2) conditions;\cite{Evangelista20_104018} in that work, v2RDM energies were shown to deviate from full CI ones by 10-50 mE$_{\rm h}$ for 2- and 3-dimensional clusters at stretched geometries. Here, we have expanded the benchmark study of Ref.~\citenum{Evangelista20_104018} to include v2RDM calculations performed under full three-particle (3POS) conditions. 3POS consistently improves upon PQG+T2 results, providing quantitatively-accurate energies in the case of the 1-dimensional (chain and ring) H$_{10}$ models. As for the 2- and 3-dimensional geometries, 3POS does reduce the energy errors exhibited under the PQG+T2 conditions, but energy errors exceeding 20 mE$_{\rm h}$ are still observed at stretched geometries. We have found that these sizeable errors are often associated with large   3-body correlations, as measured by $||{}^3\lambda||^2 / ||{}^2\lambda||^2$, but this metric is not a universal indicator of the reliability of 3POS. First, in the case of the hydrogen chains, large values of $||{}^3\lambda||^2 / ||{}^2\lambda||^2$ indicate potentially problematic cases, but small values do not necessarily guarantee reliable energetics. Second, in molecular systems such as carbon monoxide, values of $||{}^3\lambda||^2 / ||{}^2\lambda||^2$ exceeding 0.4 are observed at stretched geometries where 3POS reproduces CI-CASSCF energies to within only a few tenths of one mE$_{\rm h}$.

Despite the critical lens through which we view v2RDM results for the 2- and 3-dimensional hydrogen cluster models, the energies obtained under 3POS conditions are actually quite accurate, when compared to those obtained using other approximate correlation models. As shown in Ref.~\citenum{Evangelista20_104018}, standard low-order coupled-cluster methods ({\em i.e.}, coupled cluster (CC) with single and double excitations [CCSD],\cite{Bartlett82_1910} CCSD with perturbative triple excitations [CCSD(T)],\cite{HeadGordon89_479}  and completely renormalized CC with perturbative triple excitations [CR-CC(2,3)]\cite{Wloch05_224105}) all display larger errors than v2RDM with PQG+T2 conditions, for all four of the H$_{10}$ model systems. CC methods tend to diverge at larger $r$, while v2RDM calculations display comparatively moderate errors in this limit. Hence, the v2RDM approach can be considered a mostly reliable one for strong correlation problems, particularly when the optimal RDMs satisfy full three-particle $N$-representability conditions.  Nonetheless, v2RDM-derived results should be viewed with caution when large values of $||{}^3\lambda||^2 / ||{}^2\lambda||^2$ are observed.

\vspace{0.5cm}

{\bf Supporting Information} Full CI and v2RDM (PQG, PQG+T2, and 3POS) energies for the H$_{10}$ clusters considered in this work; CI- and v2RDM- (3POS) driven CASSCF energies for N$_2$ and CO dissociation curves.

\vspace{0.5cm}

{\bf Acknowledgments} This material is based upon work supported by the Army Research Office Small Business Technology Transfer (STTR) program under Grant No. W911NF-19-C0048.

\noindent {\bf DATA AVAILABILITY}\\

    The data that support the findings of this study are available from the corresponding author upon reasonable request.

\bibliography{3pos.bib}

\end{document}